\documentclass[runningheads]{llncs}
\usepackage[T1]{fontenc}
\usepackage{graphicx}
\usepackage{multicol}
\usepackage{multirow}
\usepackage{booktabs}

\usepackage{float}
\usepackage{amsmath}
\usepackage{amssymb}
\usepackage{svg}
\usepackage{caption}
\usepackage{orcidlink}
\usepackage{marvosym}

\begin{document}
\title{SGSR: Structure-Guided Multi-Contrast MRI Super-Resolution via Spatio-Frequency Co-Query Attention}
%

%
\author{Shaoming Zheng\orcidlink{0000-0001-5628-4311}, Yinsong Wang\orcidlink{0009-0008-7288-4227}, Siyi Du\orcidlink{0000-0002-9961-4533}, \and
Chen Qin\textsuperscript{(\Letter)}\orcidlink{0000-0003-3417-3092}}
\authorrunning{S. Zheng et al.}
\titlerunning{Structure-Guided Multi-Contrast MRI Super-Resolution}
%
\institute{Department of Electrical and Electronic Engineering \& I-X, Imperial College London, London, UK\\
\email{\{s.zheng22,y.wang23,s.du23,c.qin15\}@imperial.ac.uk}}

\maketitle              
\begin{abstract}
Magnetic Resonance Imaging (MRI) is a leading diagnostic modality for a wide range of exams, where multiple contrast images are often acquired for characterizing different tissues. 
However, acquiring high-resolution MRI typically extends scan time, which can introduce motion artifacts. Super-resolution of MRI therefore emerges as a promising approach to mitigate these challenges.
Earlier studies have investigated the use of multiple contrasts for MRI super-resolution (MCSR),
whereas majority of them did not fully exploit the rich contrast-invariant structural information.
To fully utilize such crucial prior knowledge of multi-contrast MRI, in this work, we propose a novel structure-guided MCSR (SGSR) framework based on a new spatio-frequency co-query attention (CQA) mechanism. Specifically, CQA performs attention on features of multiple contrasts with a shared structural query, which is particularly designed to extract, fuse, and refine the common structures from different contrasts. We further propose a novel frequency-domain CQA module in addition to the spatial domain,  to enable more fine-grained structural refinement. Extensive experiments on fastMRI knee data and low-field brain MRI show that SGSR outperforms state-of-the-art MCSR methods with statistical significance.

\end{abstract}


\section{Introduction}

Magnetic Resonance Imaging (MRI) is one of the most useful and important imaging techniques in hospitals for the diagnosis of diseases, due to its safety without ionizing radiation and its superior soft tissue contrast. However, the acquisition of high-resolution (HR) MRI is often time-consuming, and thus acquiring multiple images of that with different contrasts becomes even more challenging. To mitigate the challenges, super-resolution (SR) of MRI has emerged as a promising approach to improve the spatial resolution of images and restore tissue contrast.
Since multi-contrast images in MRI provide abundant and complementary information for characterizing tissues, they can be jointly exploited in multi-contrast SR (MCSR) for enhancing the SR performance. In principal, multi-contrast images of the same subject should share contrast-invariant information that corresponds to the underlying spatial structure such as object edges, whereas they also possess individual contrast-specific attribute that captures the rendering of structure determined by imaging physics, e.g., tissue contrast. Exploring such attributes in multi-contrast MRI is therefore assumed to be able to provide effective and complementary knowledge for the MCSR task. 

Unlike most of the early SR models that mainly focus on single image SR (SISR) without leveraging complementary information within multiple correlated images, recent works have considered the multi-image SR via utilizing a reference (Ref) image as guidance to super-resolve low resolution (LR) images. Examples of those include TTSR~\cite{yang_learning_2020} and MASA~\cite{lu_masa-sr_2021}, which used the Ref image as a textural guidance for SR similar to style transfer tasks, without specifically considering the structural information across images, and thus could result in potential lack of structural coherence. Recent studies in MRI SR have also started to leverage the availability of multi-contrast images, where they similarly also utilize an available high-resolution contrast image as the Ref.
Under this setup, Lyu et al~\cite{lyu2020multi}, MINet~\cite{feng_multi-contrast_2021} and SANet~\cite{feng_exploring_2024} have been proposed and demonstrated the superiority of MCSR over SISR. They mainly used relatively simple contrast fusion mechanisms such as feature concatenation and separable attention for leveraging multi-contrast information. To improve upon that, DCAMSR~\cite{huang_accurate_2023} has introduced spatial-channel attention to better capture cross-contrast semantics. Besides, various works~\cite{sun_deep_2019,zhou_dudornet_2020,lyu_dudocaf_2022,guo_reconformer_2024,feng_multimodal_2023,zhao_jojonet_2022} have also proposed multi-contrast fusion mechanisms for MRI reconstruction. However, they are all designed to equally attend to each region without explicitly focusing on structural parts. Therefore they cannot distinguish structural and non-structural features, which can thus be inefficient in exploiting the cross-contrast information for MCSR. 

To overcome this, there have been some efforts in investigating the explicit incorporation of structural information in MCSR. For instance, 
McMRSR++~\cite{li_rethinking_2023} have used the high-frequency signals (commonly representing the structural information) of the HR Ref as an additional target for the SR, which enables the model to be structure-aware by enforcing it to recover the structural target from the latent space. 
WavTrans~\cite{li_wavtrans_2022} similarly proposed to perform contrast fusion on the high-frequency wavelet components of the Ref to fully exploit the underlying structural knowledge. However, the structural information assumed in their works are extracted either from Ref or LR contrast, without exploiting the complementary structural information among them. 

In this work, we propose SGSR, a novel super-resolution framework that leverages the prior knowledge in multi-contrast images for MCSR via exploiting contrast-invariant structural information. In particular, we propose a co-query attention (CQA) mechanism driven by the shared-structure characteristic in multi-contrast images, where contrast-invariant features (i.e., structures) are leveraged jointly from multiple contrasts, constituting the `query', and contrast-specific characteristics (i.e., appearances) are distilled and interacted with structural information in individual contrast respectively. 
We further propose to perform CQA in both spatial and frequency domains to exploit their complementary mechanisms in representing and processing signals. In particular, the spatial CQA can attend to local features while the frequency CQA enables more fine-grained structure-appearance interactions. We observe that CQA is more parameter-efficient compared to cross-attention methods. Our experiments on knee and brain MRI datasets showcase that SGSR outperforms the state-of-the-art MRI SISR and MCSR methods both quantitatively and qualitatively. To clarify our novelty, we are the first to explicitly exploit contrast-invariant structures with attention mechanisms in MCSR.

\section{Methodology}




\begin{figure}[t!]
    \centering
    \includegraphics[width=1\linewidth]{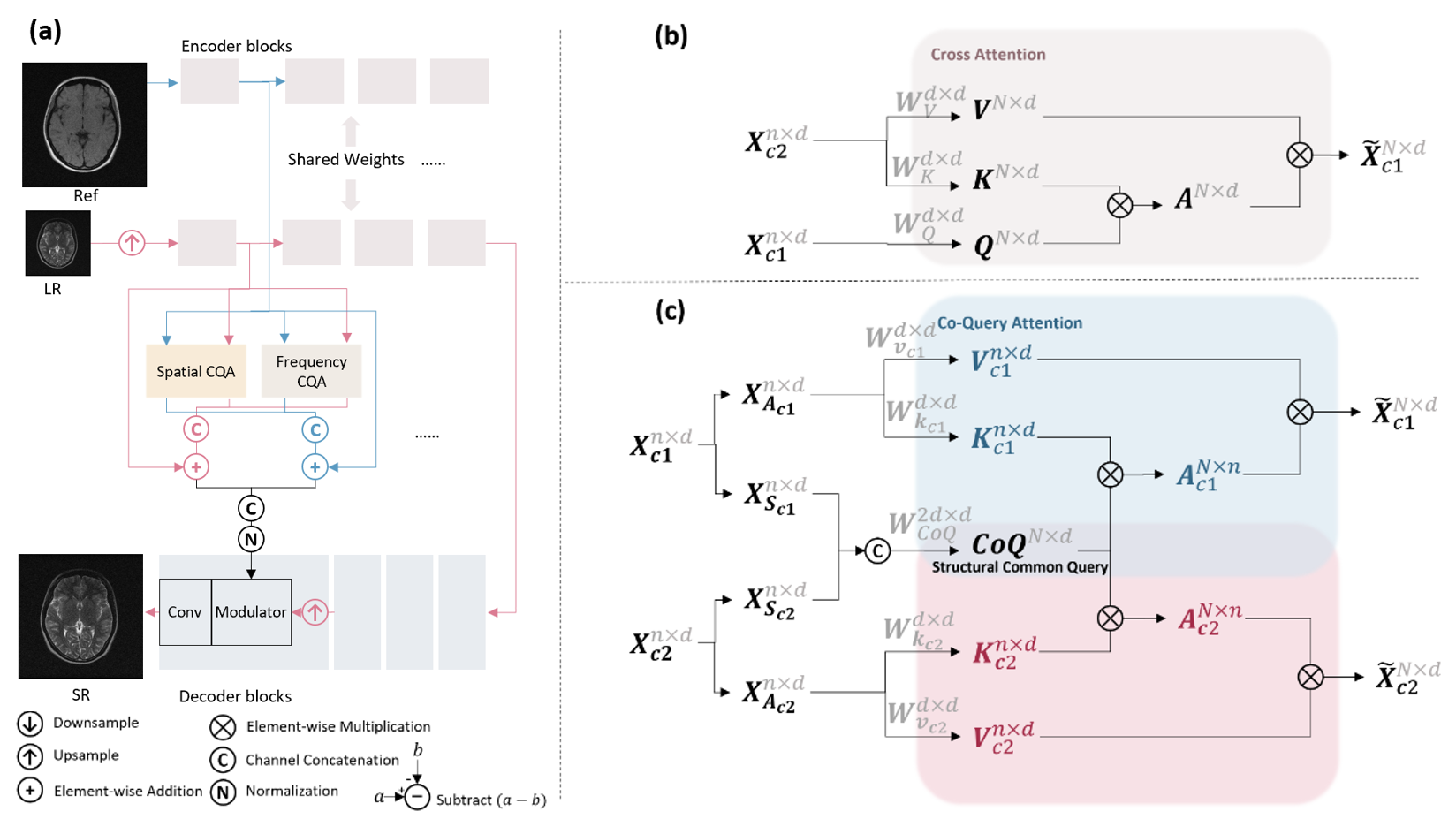}
    \caption{(a) Overall architecture of our SGSR; (b) Standard cross-attention; (c) The proposed co-query attention (CQA) within the spatial module.}
    \label{fig:arch}
\end{figure}

Our proposed SGSR consists of three components as shown in Fig. \ref{fig:arch}(a). First, the spatial co-query attention (SCQA) module extracts and refines common structures from multiple contrasts via an efficient spatial attention between structures and appearances (Sec. 2.1). Second, the frequency co-query attention (FCQA) module performs CQA in parallel to the SCQA, formulating token as frequency components to further enable more fine-grained of structure-appearance interactions (Sec. 2.2). Finally, an encoder-decoder network serves as the backbone, enabling the attention learning on the feature pyramid of the image. (Sec. 2.3).

\subsection{Spatial Co-Query Attention (SCQA)} 
\noindent\textbf{Co-Query Attention (CQA):} 
Conventionally, cross-attention can be used for dealing with information transfer between components, where the attention is performed between queries of one contrast and keys and values of another one, as shown in Fig.\ref{fig:arch}(b). However, in MSCR, as we discussed earlier, it is desirable to exploit common contrast-invariant information among multiple contrasts, whereas the cross attention mechanism is limited in this aspect due to its inherent design.  To tackle this issue, we therefore propose the query-sharing design of CQA. 
Specifically, as shown in Fig. \ref{fig:arch}(c), CQA is composed of a shared `query' representing the common structures among contrasts and individual `keys' and `values' corresponding to contrast-specific features. The attention between keys and query therefore reflects the structure-appearance interactions within images, exploring structure-related contrast information. 


In detail,  we assume that multi-contrast images can be extracted into structural features $X_{S_c}$ of size $N\times d$ and appearance features $X_{A_c}$ of size $n\times d$, which serve as the input of CQA. Here $c$ represents different contrasts, $N$ and $n$ are the number of tokens, and $d$ is the number of feature channels. As shown in Fig. \ref{fig:arch}(c), CQA first concatenates the input structural features $X_{S_c}$ of each contrast along the channel dimension and map it from $2d$ to $d$-dimension to generate the co-query $CoQ$ with size of $N\times d$. Then, for each contrast $c$, it further generates contrast-specific keys $K_c$ and values $V_c$ with size of $n\times d$ from the appearances features $X_{A_c}$.
The CQA attention is then computed between the shared query and each of the contrast-specific keys and values, which can be formulated as:

\begin{equation}
    \tilde{X}_c = \sigma\left(\frac{Q (K_c)^T}{\sqrt{d}}\right)V_c.
\end{equation}
\noindent Here $\tilde{X}_c$ denote the output of CQA, which can be interpreted as structural features refined with the appearances of contrast $c$. Note that unlike the single attention output as in cross-attention, CQA produces output for each contrast.

\noindent\textbf{CQA in Spatial Domain:} 
We first propose to adapt CQA in the spatial domain to enable interactions among local image features. For extracting structural and appearance features as inputs to the SCQA, we propose to adopt a simple strategy of low pass filtering to approximate them based on the assumption that low frequency components contain most of the information of image contrast while high frequency signals correspond to information about the structures, e.g., edges and boundaries of the image \cite{du_adaptive_2021}. In detail, the appearance features $X_{A_c}$ in spatial domain can be obtained via a downsampling operation on the encoder features $X_c$, and the structural features $X_{S_c}$ can be derived by the subtraction of the upsampled $X_{A_c}$ from $X_c$. Given these, the computational complexity of the CQA in the spatial module would be $O(Nnd)$ (typically, $n \ll N$) instead of the quadratic $O(N^2d)$ in vanilla cross-attention, which is therefore more efficient.

\subsection{Frequency Co-Query Attention (FCQA)} 


\begin{figure}[t!]
    \centering
    \includegraphics[width=1\linewidth]{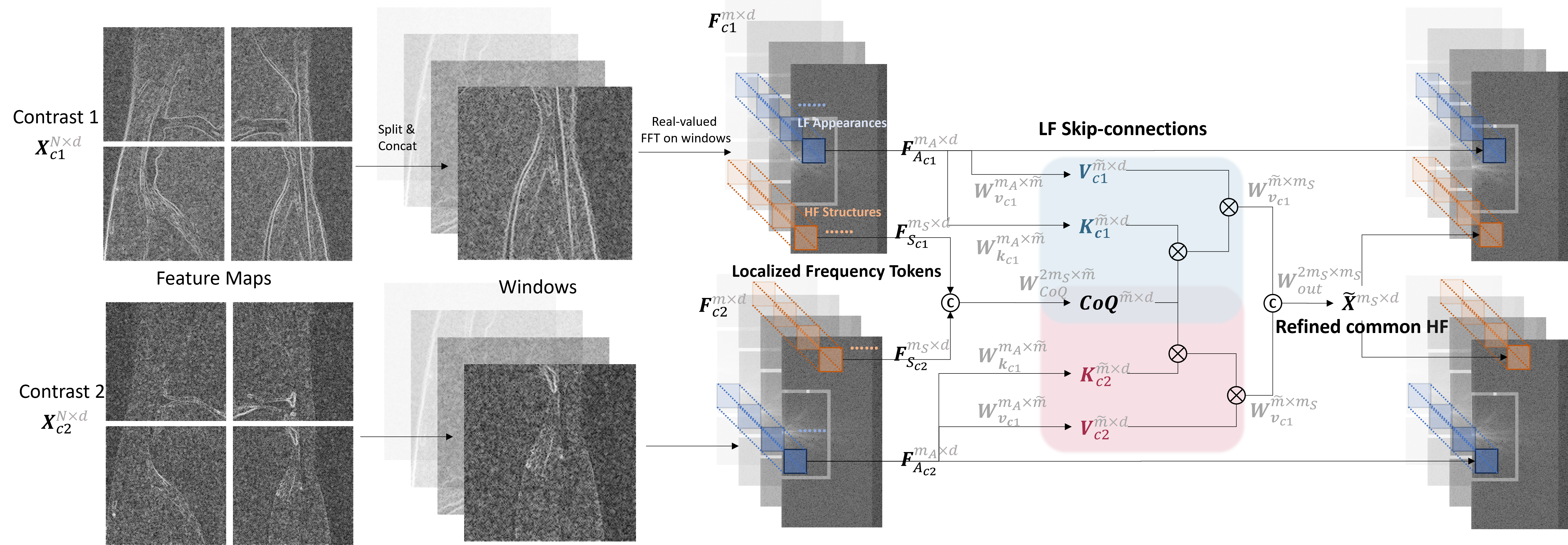}
    \caption{Frequency co-query attention (FCQA) module. The above process is followed by Inverse FFT to transform back to spatial domain.}
    \label{fig:fcqa}
\end{figure}

 To fully exploit the common structural knowledge and further refine it beyond local spatial features, we propose to extend CQA to the 2D Fourier frequency domain, named Frequency Co-Query Attention (FCQA). In contrast to SCQA, each token in FCQA represents a frequency component instead of a spatial position as shown in Fig. \ref{fig:fcqa}. This extends the bi-partition of high-frequency structures and low-frequency appearances in SCQA to more fine-grained frequency levels,  which therefore allows the exploitation of finer structure-appearance relationships. In the FCQA module, it consists of a transformation from spatial features to localized frequency tokens, a CQA module in the frequency domain, and an inverse transformation back to the spatial domain.

\noindent\textbf{Localized Frequency Tokens:} To transform from spatial to the frequency domain, the FCQA module first performs 2D Fast Fourier Transform (FFT) within local windows of the encoder feature maps $X_c$. This aims to reduce the number of frequency components and thus the attention tokens compared with global FFT for efficient computation.  
We perform real-valued FFT on each window as it can halve the number of resulting frequency components (denoted as $m$) compared to complex FFT. 
Then for each 2D spatial frequency component, we concatenate its $d$-dimensional tokens in each window
to form a final token of dimension $n_f d$, where $n_f$ is the number of windows. Our final frequency domain $F_c$ is therefore composed of $m$ such localized frequency tokens. If we use these tokens for attention, the computational complexity would be $O(m^2 \cdot n_fd) = O((N/n_f)^2 \cdot n_fd) = O(N^2d/n_f)$, which is also sub-quadratic as in SCQA. In essence, we found that concatenating windows along the channel significantly improve the efficiency.

\noindent\textbf{CQA in Fourier Frequency Domain:}
Similar to the strategy in SCQA, we separate structures- and appearances-relevant features by grouping the localized frequency tokens into high-frequency structural tokens and low-frequency appearance tokens. Specifically, we take the central $1/f$ part of $F_c$ as appearance features $F_{A_c}$ with $m_A$ tokens and the rest as structural features $F_{S_c}$ with $m_S$ tokens, as shown in Fig. \ref{fig:fcqa}, where $f$ is the hyperparameter for frequency splitting. 
To achieve efficient computation and utilize the sparsity of frequency domain, we further reduce the number of both structural and appearance tokens to $\tilde{m}$ via linear projections. In this way, the computational complexity of attention can be improved to $O(\tilde{m}^2 d) < O(N^2d/n_f)$. We then merge the attention result of each contrast with the original appearance tokens as shown in the ``LF skip-connections'' in figure \ref{fig:fcqa} to restore the size of the original frequency domain $F_c$. Then we perform inverse FFT to transform the refined structural features back to spatial domain for downstream combination with spatial features.

\subsection{Super-resolution Backbone}
We use an encoder-decoder network with 4 residual groups of convolutional layers as the base of our overall architecture with default configurations as in \cite{huang_accurate_2023}. The encoder backbone encodes images of all contrasts into multi-scale spatial features $X_c$, which are provided to both SCQA and FCQA modules. 
In the decoder, we gradually recover the latent features of LR contrast to the pixel space by using the latent features to modulate the refined structural features from the CQA modules, similar to \cite{huang_accurate_2023}. 
In this way, the upsampling process can obtain structural guidance at each scale. The overall architecture of the backbone is shown in Fig. \ref{fig:arch}(a). The model is then trained with an L1 loss between the SR and the HR.



\section{Experiments}

\noindent\textbf{Experimental Setup.}
We adopted two datasets: (1) fastMRI knee~\cite{zbontar_fastmri_2019}, with PDFS contrast as LR and PD contrast as Ref. We used 227 train/validation and 24 test pairs. (2) M4Raw~\cite{lyu_m4raw_2023}, a dataset of 0.3T low-field brain MRI from 183 participants, where each has T1, T2 and FLAIR contrasts. We use T1 images as Ref and T2 for SR. Following~\cite{huang_accurate_2023}, we used 128 train/validation pairs and 30 testing pairs. To generate the $s\times$ downsampled LR images ($s\in\{2, 4\}$), we keep the $1/s^2$ values in the central square window of k-space of the images, followed by 2D inverse FFT to convert them back into image domain.

\noindent\textbf{Implementation Details.}
We used Adam optimizer with a batch size of 4 on 4 NVidia A5000 GPUs. The initial learning rate for SANet~\cite{feng_exploring_2024} was set to $4\times10^{-5}$ according to~\cite{huang_accurate_2023}, and $2\times10^{-4}$ for the other methods. The learning rate was decayed by a factor of 0.1 starting from the 40\textsuperscript{th} epoch. The performance was evaluated for $2\times$ and $4\times$ SR in terms of PNSR and SSIM. In SCQA, we downsample the appearances features to $16\times16$. We set $f=4$ in FCQA.

\noindent\textbf{Quantitative Comparison Study.}
We compare our SGSR with representative methods in diverse categories: (1) SISR methods: SwinIR~\cite{liang_swinir_2021} and ELAN~\cite{zhang_efficient_2022}, (2) reference-guided SR methods: TTSR~\cite{yang_learning_2020} and MASA~\cite{lu_masa-sr_2021}, and (3) state-of-the-art methods specifically targeting MRI MCSR: SANet~\cite{feng_exploring_2024} and DCAMSR~\cite{huang_accurate_2023}. The results are summarized in Table 1. Our SGSR achieves the best performance across all metrics, scale factors, and datasets for both SISR and MCSR. For MCSR, SGSR achieves 0.1dB improvements in PSNR with 57\% less model parameters than the SOTA (DCAMSR) in M4Raw 4$\times$ SR, validating the representation power and low parameter cost of designing inductive bias for structural refinement in SGSR. For fairer comparison, when the number of parameters of DCAMSR~\cite{huang_accurate_2023} matches ours at 4.0M by reducing network complexity, ours achieves even much higher performance gain over it (+0.78dB for M4Raw 4$\times$, +0.41dB for fastMRI 4$\times$, +0.32dB for M4Raw 2$\times$, +0.1dB for fastMRI 2$\times$). As shown in Table \ref{table:efficiency}, the attention of our modules are superior in all aspects (10$\times$ less FLOPs and 5-10$\times$ less memory), aligning with our complexity analysis in methodology. Compared with general reference-guided SR methods, SGSR shows even better improvements (+1.1dB in PSNR over MASA) as these methods do not explicitly attend to the structural context due to their limited receptive field and weak fusion mechanisms designed for local textures. We achieve similar gains on fastMRI 4$\times$ compared to how DCAMSR improves on MASA (second best) (+0.2dB). SGSR also generalizes well to SISR. In the SISR experiment of our SGSR, we use the copy the LR as Ref. In this way, SGSR tends to extract and refine the structures within the LR contrast and thus achieves better quality than both global attention (ELAN) and local attention (SwinIR) methods without any structure-exploiting designs.

\begin{table}[t!]
\setlength{\tabcolsep}{1.5 mm}
\caption{Quantitative comparisons. * stands for values that our method significantly outperforms with $p$-value $< 0.01$. MASA-SR and TTSR are unable to perform 2× SR based on their official implementation.}
\centering
\resizebox{\textwidth}{!}{\begin{tabular}{llllllllll}
  \toprule[1pt]
	\multirow{3}{*}{Methods} & \multicolumn{4}{c}{fastMRI (knee)} & \multicolumn{4}{c}{M4Raw (brain)} & \\
 
   \cmidrule(lr){2-9}
	
    & \multicolumn{2}{c}{$2\times$} & \multicolumn{2}{c}{$4\times$} & \multicolumn{2}{c}{$2\times$} & \multicolumn{2}{c}{$4\times$} & \\
 
   \cmidrule(lr){2-9}
	& PSNR$\uparrow$ & SSIM$\uparrow$ & PSNR$\uparrow$ & SSIM $\uparrow$ & PSNR$\uparrow$ & SSIM$\uparrow$ & PSNR$\uparrow$ & SSIM $\uparrow$ & \#Params$\downarrow$ \\
	\cmidrule(lr){1-10} 
    SwinIR (SISR) & 32.01* & 0.715* & 30.73* & 0.628* & 32.16* & 0.777* & 29.73* & 0.709* & 11.9M \\
    ELAN (SISR) & 32.03* & 0.715* & 30.42* & 0.618* & 31.71* & 0.770* & 28.72* & 0.680* & 8.3M \\
	DCAMSR (SISR) & 32.07* & 0.717* & 30.71* & 0.627* & 32.19* & 0.777* & 29.74* & 0.709* & 9.3M \\
    \textbf{SGSR (SISR)} & \textbf{32.12} & \textbf{0.719} & \textbf{30.86} & \textbf{0.632} & \textbf{32.22} & \textbf{0.778} & \textbf{29.80} & \textbf{0.713} & \textbf{4.0M} \\
    \cmidrule(lr){1-10}
    MASA-SR & - & - & 30.77* & 0.628* & - & - & 29.48* & 0.703* & 4.03M \\
    TTSR & - & - & 30.64* & 0.627* & - & - & 29.86* & 0.712* & 6.42M \\
    SANet & 32.00* & 0.716* & 30.41* & 0.622* & 32.06* & 0.775* & 29.47* & 0.704* & 11M \\
    DCAMSR & 32.20* & 0.721* & 30.97* & 0.637* & 32.31* & 0.779* & 30.48* & 0.728* & 9.3M \\
    DCAMSR (4M) & 32.13* & 0.720* & 30.64* & 0.626* & 32.07* & 0.775* & 29.80* & 0.712* & 4.0M \\
    \textbf{SGSR (Ours)} & \textbf{32.23} & \textbf{0.723} & \textbf{31.05} & \textbf{0.640} & \textbf{32.39} & \textbf{0.780} & \textbf{30.58} & \textbf{0.731} & \textbf{4.0M} \\
	\bottomrule[1pt]
\end{tabular}}
\label{table:quant}
\end{table}

\begin{figure}[t!]
    \centering
    \includegraphics[width=1\linewidth]{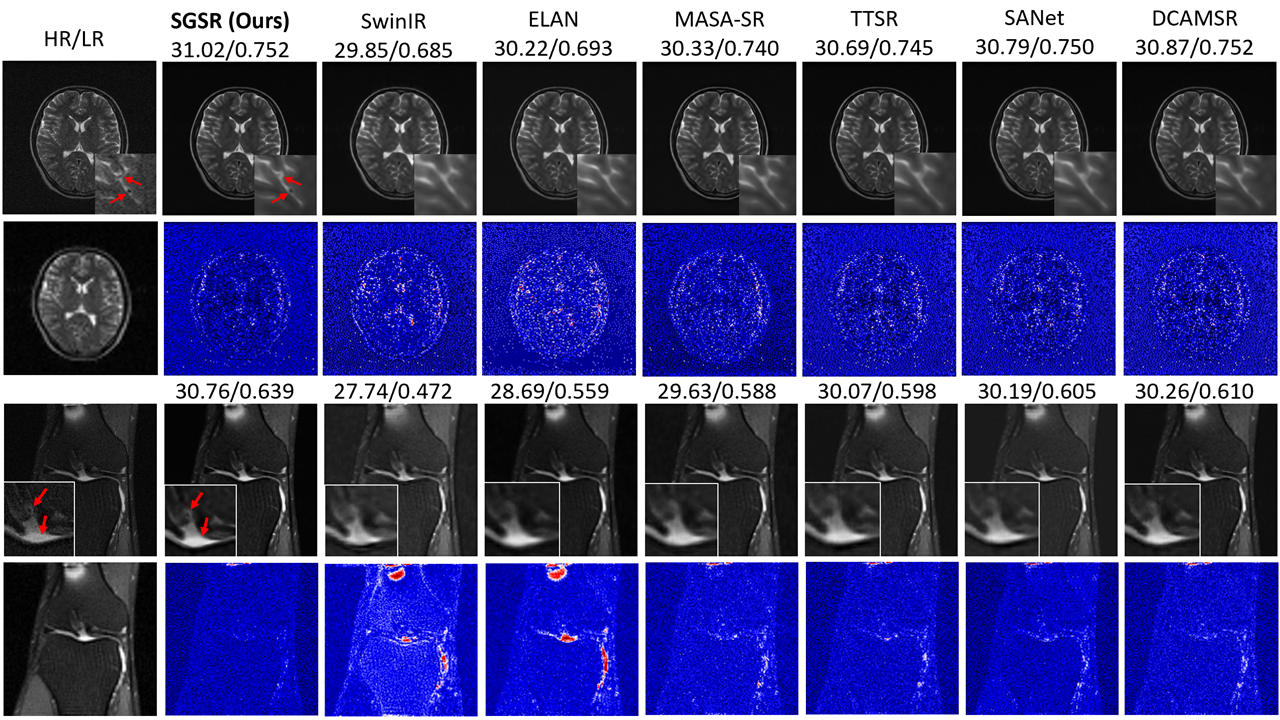}
    \caption{Qualitative results on SR predictions and error maps of fastMRI ~\cite{zbontar_fastmri_2019} (top) and M4Raw\cite{lyu_m4raw_2023} (bottom). Metrics are shown in PSNR/SSIM for each image. Input (bottom) and ground truth (top) are shown in the first column.}
    \label{fig:error}
\end{figure}

\noindent\textbf{Qualitative Results.}
Visual comparisons are shown in Fig. \ref{fig:error} for 
$4\times$ SR. 
Our method produces overall the least errors and accurately recovers very small structures such as those indicated by arrows. We attribute this to our model's capacity to refine structures leveraging different contrasts compared to other methods.

\noindent\textbf{Ablation Study.}
We conducted ablation studies on the M4Raw\cite{lyu_m4raw_2023} $4\times$ SR as shown in Table \ref{table:ablation}. Four variations of our SGSR are experimented: (1) \textit{w/o Ref} (i.e. SISR): we replace Ref input by upsampled LR. (2) \textit{CA} (cross-attention): i.e. DCAMSR\cite{huang_accurate_2023}. (3) \textit{FCQA}, and (4) \textit{SCQA}. The gain of SGSR (\textit{FCQA+SCQA}) from \textit{CA} validates the superiority of exploiting common structural representation which undergoes refinement with each contrast. In addition, the gains of both \textit{SCQA} and \textit{FCQA} from \textit{CA} validate the effectiveness of integrating complementary mechanisms of structural fusion and refinement in both spatial and frequency domains. Their gain account for 70\% and 50\% of the gap between DCAMSR and our SGSR ($p\ll 0.01$). We attribute this to their respective advantages: SCQA attends to local features while FCQA enables fine-grained structure-appearance interactions. We visualize the structural queries in SCQA in the supplementary materials.

\begin{figure}
\begin{minipage}[t]{.53\textwidth}
    \captionof{table}{Ablation study on M4Raw\cite{lyu_m4raw_2023} 4× task. * for values that the full version significantly outperforms with $p$-value $< 0.01$.}
    \resizebox{\textwidth}{!}{
    \begin{tabular}{lllllll}
        \toprule[1pt]
        Variants & Ref & CA & SCQA & FCQA & PSNR & SSIM \\
        \cmidrule(lr){1-7}
        \textit{w/o Ref} & & \checkmark & \checkmark & \checkmark & 29.80* & 0.713* \\
        \textit{CA} & \checkmark & \checkmark  & & & 30.48* & 0.729* \\
        \textit{FCQA} & \checkmark & & & \checkmark & 30.53* & 0.729* \\
        \textit{SCQA} & \checkmark & & \checkmark && 30.55* & 0.729* \\
        \textbf{FCQA+SCQA} & \checkmark & & \checkmark & \checkmark & \textbf{30.58} & \textbf{0.731} \\
        \bottomrule[1pt]
    \end{tabular}}
    \label{table:ablation}
\end{minipage}
\hfill
\begin{minipage}[t]{.43\textwidth}
    \captionof{table}{Comparisons of FLOPs \& GPU memory between attention of our modules and cross attention for $128\times 128$ LR \& Ref input.}
    \resizebox{\textwidth}{!}{
    \begin{tabular}{lll}
        \toprule[1pt]
        Variants & FLOPs & Memory \\
        \cmidrule(lr){1-3}
        \textit{SCQA attention} & $2.85 \times 10^8$ & 118 MB \\
        \textit{FCQA attention} & $2.68 \times 10^8$ & 214 MB \\
        \textit{Cross attention} & $2.28 \times 10^9$ & 1030 MB \\
        \bottomrule[1pt]
    \end{tabular}}
    \label{table:efficiency}
\end{minipage}
\end{figure}

\section{Conclusion}
In this paper, we presented a novel and small structure-guided multi-contrast MRI super-resolution (SGSR) scheme. Exploiting the contrast-invariant structural information as an important prior, our method features common structural extraction and refinement from all the contrasts via a tailored co-query attention (CQA) mechanism. In addition to the CQA on spatial domain which processes local features, we further extend this attention paradigm to the frequency domain, adapting it to fine-grained structure-appearance interactions with a novel frequency co-query attention (FCQA) paradigm. Comparison study shows that our method outperforms other competing multi-contrast SR methods regarding both SR image quality and parameter cost. For future works, we will extend the method to multi-modal tasks such as super-resolution between MRI and CT.

\noindent\textbf{Acknowledgements.}
This work was partially supported by the Engineering and Physical Sciences Research Council [grant number EP/X039277/1].
%
%
%
%
\bibliography{main}
\bibliographystyle{splncs04}
\end{document}